\documentclass{article}
\usepackage{amsmath}
\usepackage{graphicx}
\usepackage{hyperref}
\usepackage{subfig}

\begin{document}
\title{Experience with GitHub Copilot for Developer Productivity at Zoominfo}
\author{Gal Bakal, Ali Dasdan, Yaniv Katz, Michael Kaufman, Guy
  Levin\footnote{The author names are in last name alphabetical order.}\\
  Zoominfo\\
  \{gal.bakal, ali.dasdan, yaniv.katz, michael.kaufman, guy.levin\}@zoominfo.com\\
}

\maketitle

\begin{abstract}
This paper presents a comprehensive evaluation of GitHub Copilot's
deployment and impact on developer productivity at Zoominfo, a leading
Go-To-Market (GTM) Intelligence Platform. We describe our systematic
four-phase approach to evaluating and deploying GitHub Copilot across
our engineering organization, involving over 400 developers. Our
analysis combines both quantitative metrics, focusing on acceptance
rates of suggestions given by GitHub Copilot and qualitative feedback
given by developers through developer satisfaction surveys. The
results show an average acceptance rate of 33\% for suggestions and
20\% for lines of code, with high developer satisfaction scores of
72\%. We also discuss language-specific performance variations,
limitations, and lessons learned from this medium-scale enterprise
deployment. Our findings contribute to the growing body of knowledge
about AI-assisted software development in enterprise settings.
\end{abstract}

\section{Introduction}
\label{sec:intro}

Developer productivity is a strategic priority at Zoominfo, driven by
our fundamental belief that the speed at which a company transforms
ideas into customer outcomes is strongly correlated with their
competitive advantage. As such, we are continuously evaluating and
adopting new methodologies and tools that enhance our developer's
productivity.

The emergence of AI-powered coding assistants has sparked significant
interest in their potential to enhance developer productivity. GitHub
Copilot (sometimes referred as ``the tool'' in this paper), launched
in 2021, represents a major advancement in this space, promising to
accelerate software development through AI-generated code
suggestions. However, while initial studies have shown promise, there
remains limited empirical evidence of its evaluation, deployment, and
effectiveness in medium- to large-scale enterprise environments.

This paper addresses this gap by presenting a detailed case study of
GitHub Copilot's production deployment at Zoominfo, a leading
Go-To-Market (GTM) Intelligence Platform, where it is used by over 400
developers who are geographically dispersed, with diverse technical
disciplines, and programming languages. Our study aims to answer five
key questions in a medium-scale enterprise setting:
\begin{enumerate}
 \item How is GitHub Copilot evaluated to reach a production
   deployment decision?
\item What are the acceptance rates for different programming
  languages?
\item What are the key factors influencing developer satisfaction with
  AI-assisted coding?
\item How effective is GitHub Copilot in improving developer
  productivity?
\item What are the observed and potential limitations of GitHub Copilot?
\end{enumerate}

We present a comprehensive case study of Zoominfo's evaluation, trial,
and company-wide deployment of GitHub Copilot. As a major GTM platform
managing hundreds of millions of business contact and company profiles
and processing billions of daily events, we believe Zoominfo's
experience offers valuable insights into the practical implementation
and impact of AI-assisted software development at scale in similar
enterprise companies.

Our analysis of GitHub Copilot usage combines both quantitative
metrics, focusing on acceptance rates of the tool's suggestions, and
qualitative feedback through developer satisfaction surveys. The
results show an average acceptance rate of 33\% for suggestions and
20\% for lines of code, with high developer satisfaction scores of
72\%. The acceptance rates we have observed are in line with the
acceptance rates reported by a few other companies in industry, e.g.,
GitHub and Google

The acceptance rates for the top four programming languages used by
our developers, i.e., TypeScript, Java, Python, and JavaScript, are
sustained at about 30\%. Interestingly we observe smaller acceptance
rates for HTML, CSS, JSON, and SQL.

Our developers primarily enjoy the time savings (around 20\%) due to
GitHub Copilot while citing as limitations the tool's lack of
domain-specific logic and lack of consistency in code quality. These
limitations negatively impact time savings due to the need for
additional scrutiny required while vetting the generated code.

Despite these limitations, we believe GitHub Copilot usage
significantly contributed to the productivity of our developers. In
addition the improvements in time savings and developer satisfaction,
the number of lines of production code contributed is on the order of
100s of 1000s of lines of code.

The rest of the paper is organized as follows. \S~\ref{sec:zi} and
\S~\ref{sec:productivity} provide the necessary background for
Zoominfo and developer productivity. \S~\ref{sec:projections} presents
the projected benefits of GitHub Copilot after an initial ad hoc
assessment. This is followed by the phases and details of our formal
evaluation in \S~\ref{sec:evaluation}. \S~\ref{sec:success} describes
how we measured success, namely, acceptance rates. The quantitative
and qualitative results due to GitHub Copilot usage in production are
discussed in \S~\ref{sec:quantitative1}, \S~\ref{sec:quantitative2},
\S~\ref{sec:quantitative3}, and \S~\ref{sec:qualitative}. The observed
and potential limitations of GitHub Copilot are presented in
\S~\ref{sec:limitations}. A fairly comprehensive presentation of
related work is in \S~\ref{sec:related}. We conclude with
\S~\ref{sec:final}.

\section{Background: Zoominfo}
\label{sec:zi}

Zoominfo\cite{Zo24} is the leading ``Go-To-Market (GTM) Intelligence
Platform''; it helps companies to acquire new customers as well as
retain and grow their existing customer base. It has comprehensive and
accurate data about 100s of millions of business contacts and
companies. It has products built on the shared platform to serve
multiple personas: Sales for sellers, Marketing for marketers, Talent
for recruiters, Operations for data managers, and Data-as-a-Service
(DaaS) for companies developing custom internal solutions.

Zoominfo has more than 37,000 customers (companies) and a few 100
thousands of monthly active users~\cite{Zo24}. It has close to 4,000
employees in total, geographically distributed over the US, Europe,
India, and Israel; a quarter of the employees are in the engineering
department, and 400 engineers are active {\em developers} producing
and deploying production software.

Zoominfo operates its platform on top of two public clouds, run by
Google and Amazon. The platform consists of many microservices and a
few monoliths interacting with data streaming and data management
platforms~\cite{ZoEn24}. The most common programming languages are
TypeScript, Python, Java, and JavaScript, distributed over 1000s of
code repositories hosted mainly with self-hosted GitHub Enterprise
Server instance but also remotely with GitLab.

The scale challenges at Zoominfo~\cite{ChSaDa24,YaAcPa24} include the
management of 100s of millions of business contact and company
profiles, keeping them accurate and up-to-date, collecting and serving
in real time billions of events per day, processing big data using AI
and serving insights to users in real-time, responding to 10s of
millions of search and other user queries a day, serving user
experience queries in less than 1 second response time, providing
99.9\% uptime per month for the most frequent user experience journeys
(which means stability objectives more stringent than 99.9\% for the
backend systems), and delivering very high security and privacy
requirements.

\section{Productivity Refresher}
\label{sec:productivity}

At Zoominfo, the productivity principle we follow is this: {\em How
  fast a company moves in transforming ideas into customer outcomes is
  the primary advantage of the company.}  We assert that companies
prioritizing rapid execution will ultimately outperform their slower
competitors, regardless of other competitive advantages those
competitors may possess. As such, developer productivity is a top
priority for us.

As detailed in \cite{Da24}, there are many factors that affect
developer productivity. Among those factors, providing the best
tooling to developers in writing, testing, and reviewing code is
fundamental. This represents our primary deployment area for GitHub
Copilot.

At a basic level productivity is defined as the amount of output
produced for a given amount of inputs. For developer productivity, the
inputs are usually limited to time spent building the output while the
output may include many features but the end result is outcomes or
value for customers.

Developer productivity is measured in two ways: Quantitative and
qualitative~\cite{Da24}. The quantitative metrics include objective
metrics like the DORA metrics~\cite{Do24a,Do24b} measured via software
development and deployment pipelines while the qualitative metrics
include subjective metrics like developer satisfaction measured via
developer surveys. Refer to \cite{Da24,FoStMa21,NoStFo23} for more
information.

\section{Projected Benefits of GitHub Copilot}
\label{sec:projections}

A few months after GitHub Copilot was released, the excitement it
generated also prompted us to conduct an ad hoc assessment of its
potential benefits. The result of this ad hoc assessment was positive,
and we proceeded with procuring GitHub Copilot for a formal
assessment.

The following benefits were projected based on this initial ad hoc
assessment.

\subsection{Augmenting Day-to-day Software Development}

\begin{itemize}

\item Automated Code Generation: GitHub Copilot can generate code
  snippets and even complete functions based on the contextual
  information provided. It can suggest logic for the developers while
  they're coding, which can be a significant time-saver, especially
  when dealing with routine or repetitive code patterns.

\item Code Review Assistance: Copilot can also serve as a
  pseudo-code-reviewer. It learns from billions of lines of code,
  meaning it can help spot potential bugs, suggest improvements, and
  ensure that the code aligns with best practices.

\item Documentation and Commenting: The AI can provide useful comments
  and assist with documentation. It can explain complex code snippets,
  making it easier for other team members to understand the codebase,
  hence promoting collaboration.

\item Learning New Technologies: When working with new languages,
  libraries, or frameworks, GitHub Copilot can be a great
  companion. It can provide code suggestions that adhere to the latest
  syntax and best practices, reducing the learning curve for
  developers.

\end{itemize}

\subsection{Overall Productivity Gains}

\begin{itemize}

\item Time Efficiency: With automated code generation and intelligent
  suggestions, developers can save significant time. This time can be
  used for more critical tasks, such as designing software
  architecture or addressing complex problems.

\item Quality Improvement: By acting as a pseudo-code-reviewer, GitHub
  Copilot can help improve the quality of the code, reducing the
  likelihood of bugs and rework.

\item Onboarding and Training: For new hires or junior developers,
  GitHub Copilot can act as a learning tool, helping them quickly
  understand the codebase, best practices, and contributing
  effectively.

\item Accelerated Development Cycles: By reducing the time spent on
  routine tasks, improving code quality, and facilitating faster
  onboarding, GitHub Copilot can significantly accelerate our
  development cycles.

\end{itemize}

\section{Methodology: From Evaluation to Rollout}
\label{sec:evaluation}

After the initial ad hoc assessment in early July of 2023, we
implemented a systematic four-phase approach from July to August of
2023 to evaluate and deploy GitHub Copilot across our engineering
organization: Initial assessment phase, trial recruitment phase, the
two-week trial phase, and the rollout phase. The rollout started in
early September of 2023 and it took a few months for GitHub Copilot to
be adopted by all the developers due to our pacing of the license
distribution.

\subsection{Phase 1: Initial Assessment Phase}

We conducted an initial qualitative assessment with five engineers
from July 10th, 2023 to July 17th, 2023 to evaluate GitHub Copilot's
potential impact on development workflows. This preliminary evaluation
focused on identifying key benefits and potential challenges within
our existing software development lifecycle.

The preliminary feedback was overwhelmingly positive, with key metrics
including:
\begin{itemize}
\item Overall experience rating: 8.8 out of 10;
\item Productivity improvement rating: 8.6 out of 10; and
\item Code standards alignment: All five participants reported good to
  excellent alignment with existing coding standards.
\end{itemize}

Qualitative feedback highlighted several key observations:
\begin{itemize}
\item Strong adaptation to existing codebase patterns and conventions;
\item No reported negative impact on code quality;
\item Minimal integration challenges with existing development
  processes; and
\item Particularly effective for unit test generation and boilerplate code.
\end{itemize}

Notable concerns included the following:
\begin{itemize}
\item Need for modification of suggested code (reported by 3 out of 5
  participants);
\item Limited visibility across multiple projects; and
\item Potential over-reliance on automated suggestions.
\end{itemize}

This initial assessment informed our subsequent trial design and
helped establish baseline metrics for the broader evaluation phase.

Procurement of GitHub Copilot for Business licenses were also acquired
during this phase to facilitate the trial and subsequent rollout.

\subsection{Phase 2: Trial Recruitment Phase}

We implemented a structured recruitment process for the controlled
trial phase, conducted from July 17th to August 14th, 2023.
The recruitment strategy employed stratified voluntary sampling to
ensure representative participation across multiple dimensions of
our engineering organization.

The trial cohort comprised 126 engineers (about 32\% of the
developers), with participation stratified across technical
specializations, experience levels, geographical locations, and
technology stacks.

To ensure compliance with organizational standards and maintain data
integrity, participants were required to fulfill several prerequisite
conditions:
\begin{enumerate}
\item Completion of internal security code review training.
\item Written acknowledgment of corporate compliance requirements:
    \begin{itemize}
    \item Generative AI usage policies,
    \item General AI governance framework,
    \item Data governance protocols,
    \item Data ethics guidelines, and
    \item Data classification standards.
    \end{itemize}
\item Commitment to provide structured feedback through a follow up
  survey.
\end{enumerate}
The public versions of some of these compliance documents can be found
at \cite{Zo24b}.

The trial protocol was designed to maintain both broad participation and 
controlled evaluation conditions while ensuring compliance with corporate 
security requirements. Participation was voluntary but managed through a 
formal application and governance framework that included:
\begin{itemize}
\item Structured application and prerequisite verification;
\item Documentation of training completion and policy acknowledgments;
\item Assignment of unique participant identifiers for tracking;
\item Comprehensive code review requirements;
\item Mandatory functionality validation and testing;
\item Documentation standards compliance;
\item Production deployment guidelines; and
\item Security compliance measures.
\end{itemize}

This comprehensive methodology enabled effective participant management
while maintaining consistent code quality standards throughout the trial
period. All participants were required to adhere to these guidelines to
ensure the validity and security of the evaluation process.

\subsection{Phase 3: Two-Week Trial Phase}

We conducted a two-week controlled trial from August 15th to August
29th, 2023, with 126 participating engineers actively integrating
GitHub Copilot into their daily development workflows.

At the trial's conclusion, we collected feedback through a
comprehensive survey (72 respondents, about 57\% response rate).

The survey addressed three key dimensions:
\begin{enumerate}
\item Overall experience and productivity impact,
\item Code quality and standards alignment, and
\item Security considerations.
\end{enumerate}

The overall experience was positive with the following ratings:
\begin{itemize}
\item Mean satisfaction rating: 8.0 out of 10, and
\item Mean productivity improvement rating: 7.6 out of 10.
\end{itemize}
Developers reported time savings for generating boilerplate and
repetitive code, unit tests, meaningful variable names, documentation,
and comments.

Regarding code quality and standards alignment, the majority of
participants were satisfied:
\begin{itemize}
\item Majority of participants reported good to excellent alignment
  with existing coding standards.
\item No participants reported a decline in code quality across their
  team's pull requests.
\item Majority of participants reported needing to make minimal
  modifications to suggested code.
\end{itemize}
Developers reported that the tool was useful across various languages
and tech stacks, although with reservations due to some
inconsistencies regarding the tool's performance on domain-specific or
highly innovative tasks, requiring additional oversight.

Regarding security considerations, survey participants demonstrated
high security consciousness:
\begin{itemize}
\item Mean security vulnerability assessment confidence: 8.2 out of 10;
\item Mean sensitive information exposure awareness: 8.2 out of 10; and
\item Mean security consideration in development: 8.6 out of 10.
\end{itemize}
Despite these high scores, developers emphasized the need for rigorous
reviews of AI-generated code to mitigate security and quality risks.

Overall, our analysis revealed strong participant satisfaction with
GitHub Copilot's core capabilities. Unit test generation and
boilerplate code creation showed the highest utility, while code
documentation and pattern recognition demonstrated moderate to high
effectiveness. Variable naming features showed modest utility.

Implementation challenges primarily involved technical integration and
context management across codebases. Despite these initial adoption
hurdles, the trial demonstrated strong positive outcomes, particularly
in security awareness and code quality maintenance, with most of
participants reporting improved productivity.

This structured trial phase was instrumental in preparing ZoomInfo for
the full rollout of GitHub Copilot, ensuring alignment with
organizational objectives and addressing key developer concerns.

\begin{figure}[ht]
  \centering
  \includegraphics[scale=0.3]{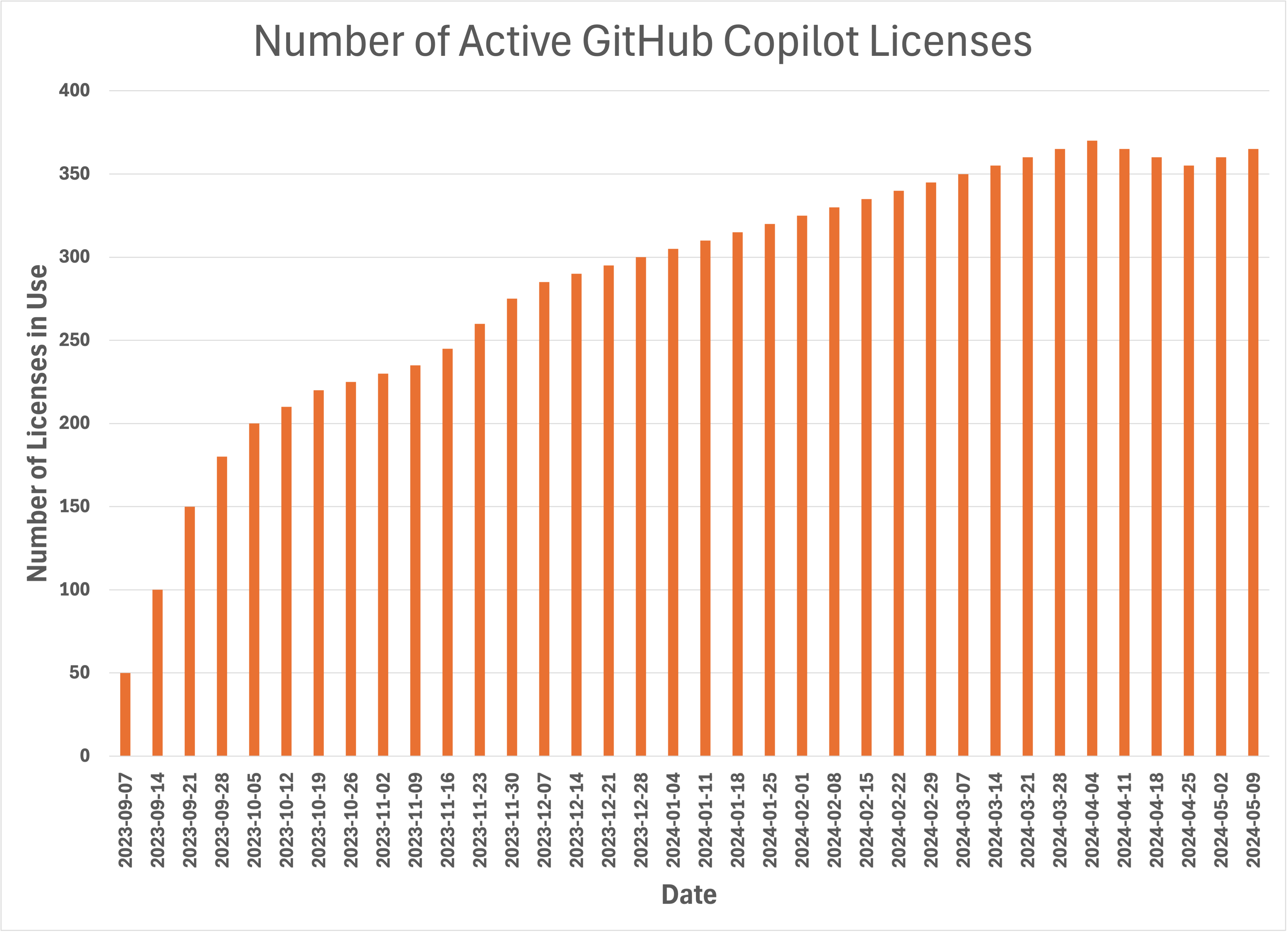}
  \caption{Adoption of GitHub Copilot during the first 8 months at
    Zoominfo. We released licenses at a controlled pace to ensure
    daily usage and success.}
  \label{fig:adoption}
\end{figure}

\subsection{Phase 4: Rollout Phase}

Following the successful trial, an analysis of the survey data from
Phase 3 determined that there were no outstanding issues that needed
to be addressed before rolling out GitHub Copilot to the remainder of
our engineers.

Thus, we initiated the full-scale deployment of GitHub Copilot across
our engineering organization on September 7th, 2023 by releasing a
ServiceNow Workflow for measured deployment of GitHub Copilot
licenses, as shown in Fig.~\ref{fig:adoption}. This streamlined the
process of GitHub Copilot provisioning and enabled tracking for
compliance and utilization metrics. This also enabled ZoomInfo
engineers to promptly get access to the tool.

\section{Success Measure}
\label{sec:success}

The impact of GitHub Copilot on developer productivity seems difficult
to measure after a short-term usage. As such, we resorted to a measure
that was recommended by GitHub after it was found to be a ``better
predictor of perceived productivity''~\cite{ZiKaLi24}: {\em Acceptance
  rate of shown suggestions}.

Acceptance rate of shown suggestions~\cite{Gi24} for one developer is
the ratio of the suggestions accepted by the developer to the total
number of suggestions shown to the developer.

For this measure, even a partial acceptance gets the full
credit. Moreover, the number of lines shown is not included in the
rate. This means a single line as well as 100s of lines get the same
credit. Both of these may be seen as limitations of this measure but
for now, the acceptance rate measure seems like a good one among the
alternatives considered in \cite{ZiKaLi24}.

The acceptance rate of shown suggestions for a team is the average
over the rates for developers.

After months of usage, we have also been monitoring for any changes in
our regular developer productivity metrics such as the DORA metrics
and developer satisfaction scores. Once we establish a reliable
causality between these metrics and the GitHub Copilot usage, we are
planning to report the results in a subsequent paper.

\begin{figure}[ht]
  \centering
  \includegraphics[scale=0.3]{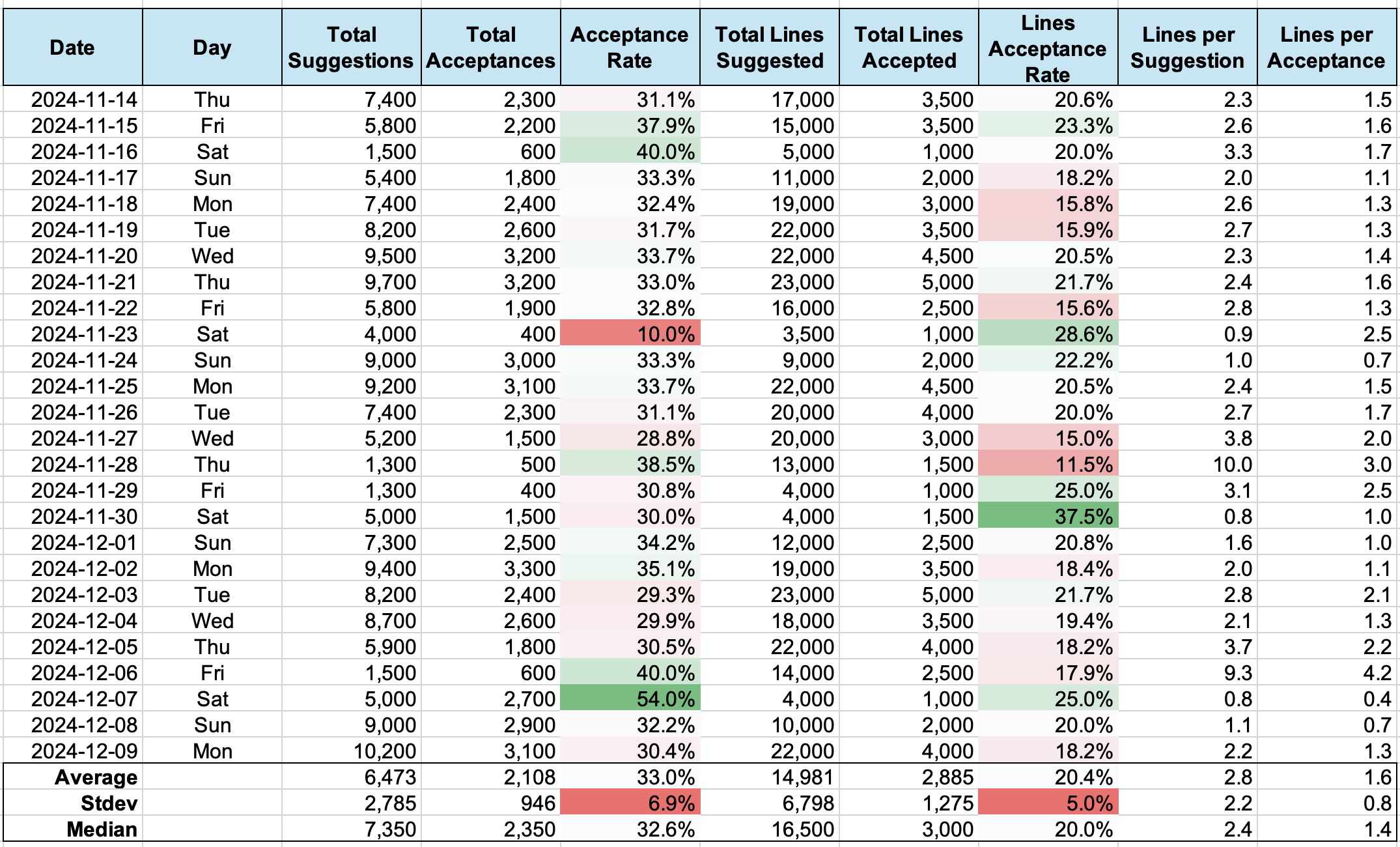}
  \caption{Daily data from Nov. 14th to Dec. 9th in 2024. For each
    day, this table shows the total number of suggestions,
    acceptances, lines suggested, lines accepted, acceptance rate,
    lines acceptance rate, lines per suggestion and lines per
    acceptance. The rows at the bottom show the averages (Average),
    standard deviations (Stdev), and medians (Median) for each
    relevant column. The conditional formatting of two columns in
    tones of green and red colors are to highlight larger values
    (green) and smaller values (red).}
  \label{fig:table}
\end{figure}

\begin{figure}%
  \centering

  \subfloat[\centering Suggestions.]{
    {\includegraphics[width=10cm]{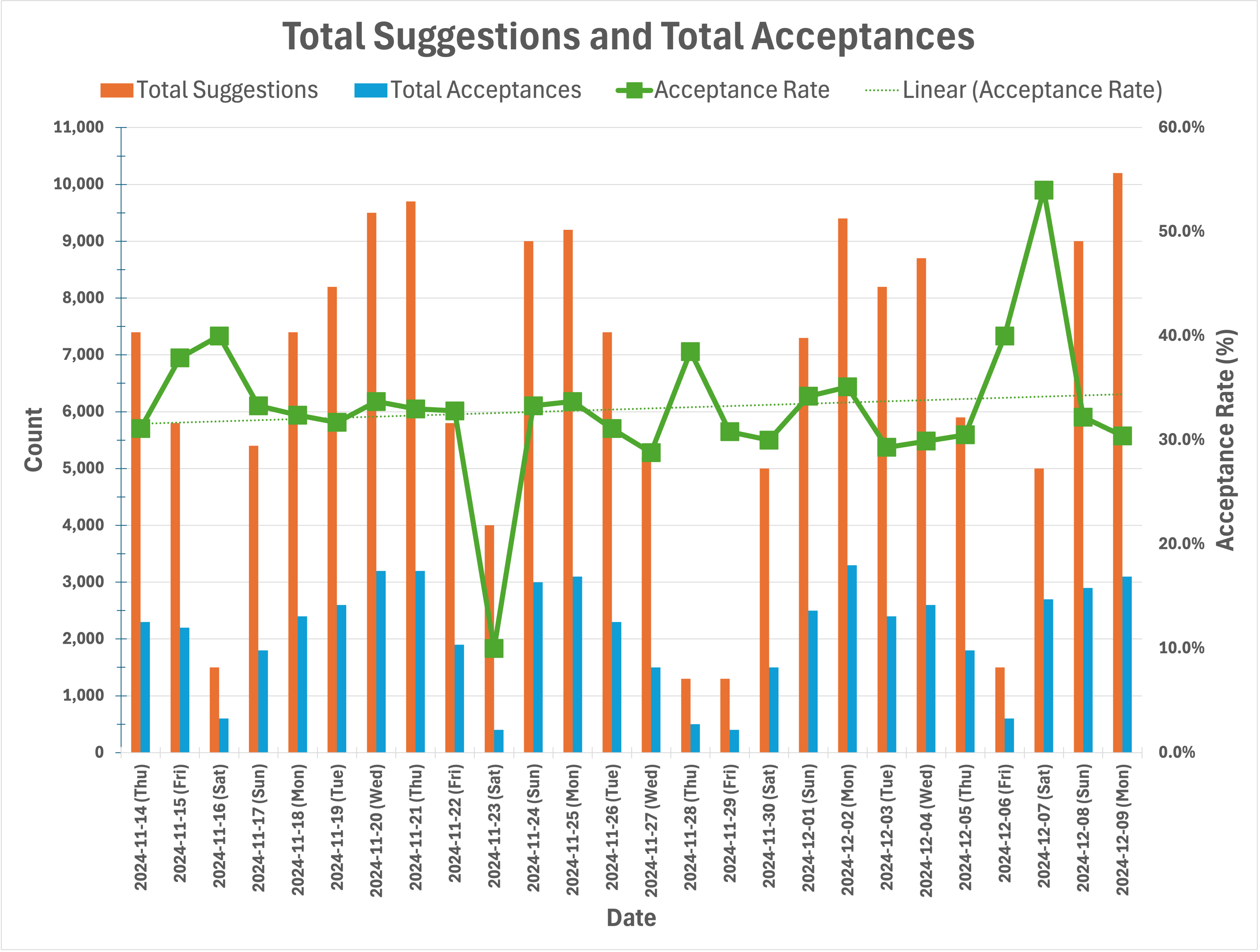} }%
  }

  \subfloat[\centering Lines.]{
    {\includegraphics[width=10cm]{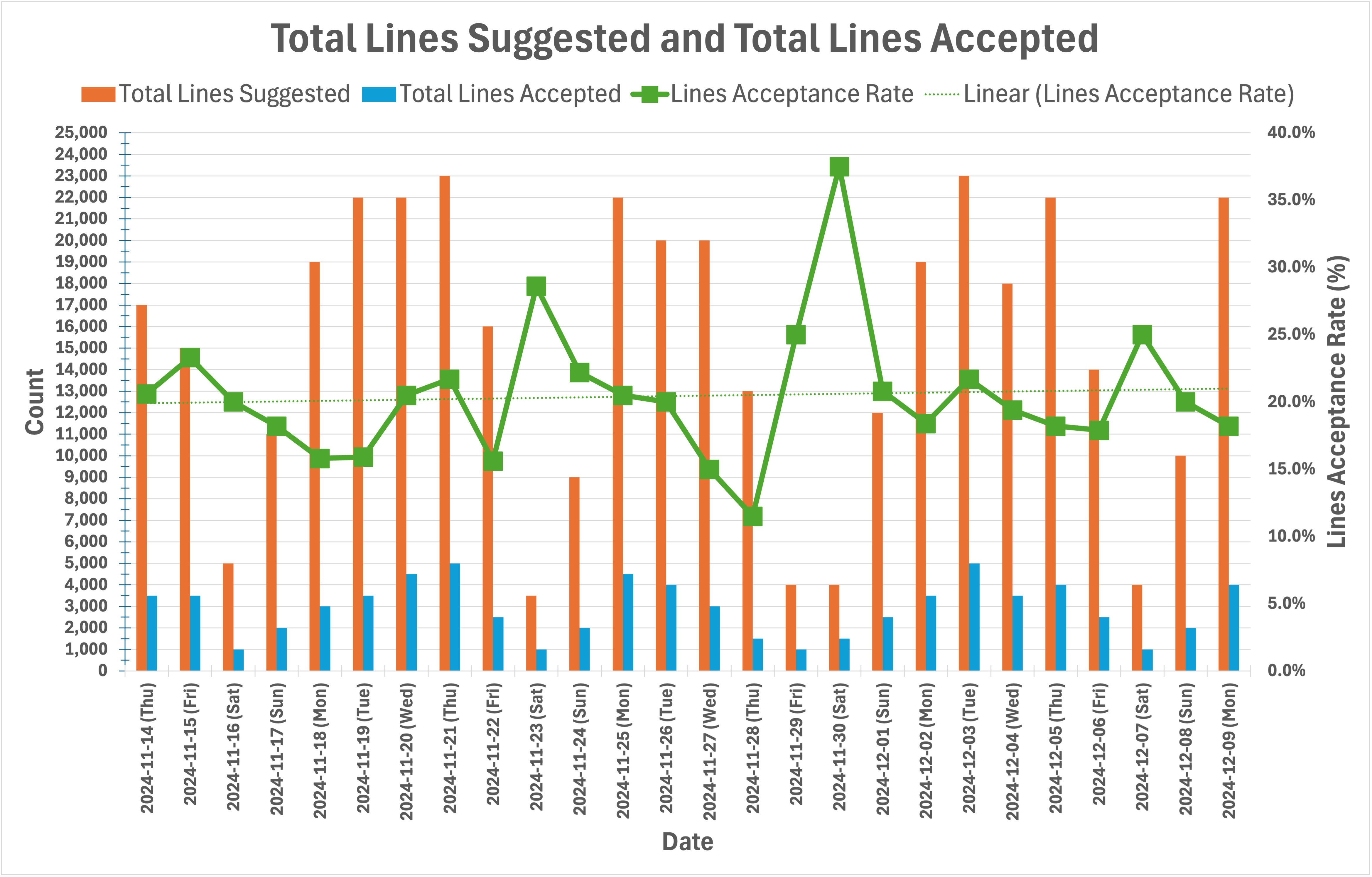} }%
  }
    
  \caption{Daily total number of suggestions and acceptances (a) and
    daily total number of suggested and accepted lines (b). Both
    figures cover the days from Nov. 14th to Dec. 9th in 2024. The
    acceptance rate in each case is the ratio of the suggested unit to
    the accepted unit. The trend lines are for the acceptance rates,
    showing a slight upward trend in each case. The wavy pattern is
    due to weekdays (high) and weekends (low).}%
  \label{fig:vertical}%
\end{figure}

\section{Quantitative Results: Overall Acceptance Rates}
\label{sec:quantitative1}

Fig.~\ref{fig:table} shows the table of data for 26 days in 2024, from
from Nov. 11th to Dec. 9th. The table shows the number of suggestions,
acceptances, lines suggested, and lines accepted.

For each prompt from a developer, GitHub Copilot makes a suggestion,
which has one or more lines of code or comments, and the developer
accepts or declines the suggestion. The ratio is the acceptance
rate. The acceptance rates in terms of suggestions and lines are also
given in the table.

The average numbers of suggestions and lines suggested are close to
6,500 and 15,000, respectively. The standard deviations are close to
the half of these numbers in each case but such large deviations are
due to the weekday and weekend periods rather than the variations from
day to day. The averages over weekdays are about 20\% and 35\% larger
than these overall averages. Similarly, the averages over weekends are
about 60\% and 75\% smaller than these overall averages.

Note that for our developers in Israel, the weekend is from Friday to
Saturday while for the developers in the US and India, the weekend is
from Saturday to Sunday.

Each suggestion has a few lines of code or comments; the average is
2.8 lines though the number of lines ranges from one to ten. The
median is also close to the average.

Fig.~\ref{fig:vertical} shows the same data in the table in
Fig.~\ref{fig:table} but as bar charts and line plots for ease of
understanding. As the acceptance rate plots show, the acceptance rates
during weekends usually increase rather than decrease. We do not know
the reason for this behavior. Also, as the acceptance rate trend lines
show, there is a slight upward trend in acceptance rates.

\begin{figure}[ht]
  \centering
  \includegraphics[scale=0.3]{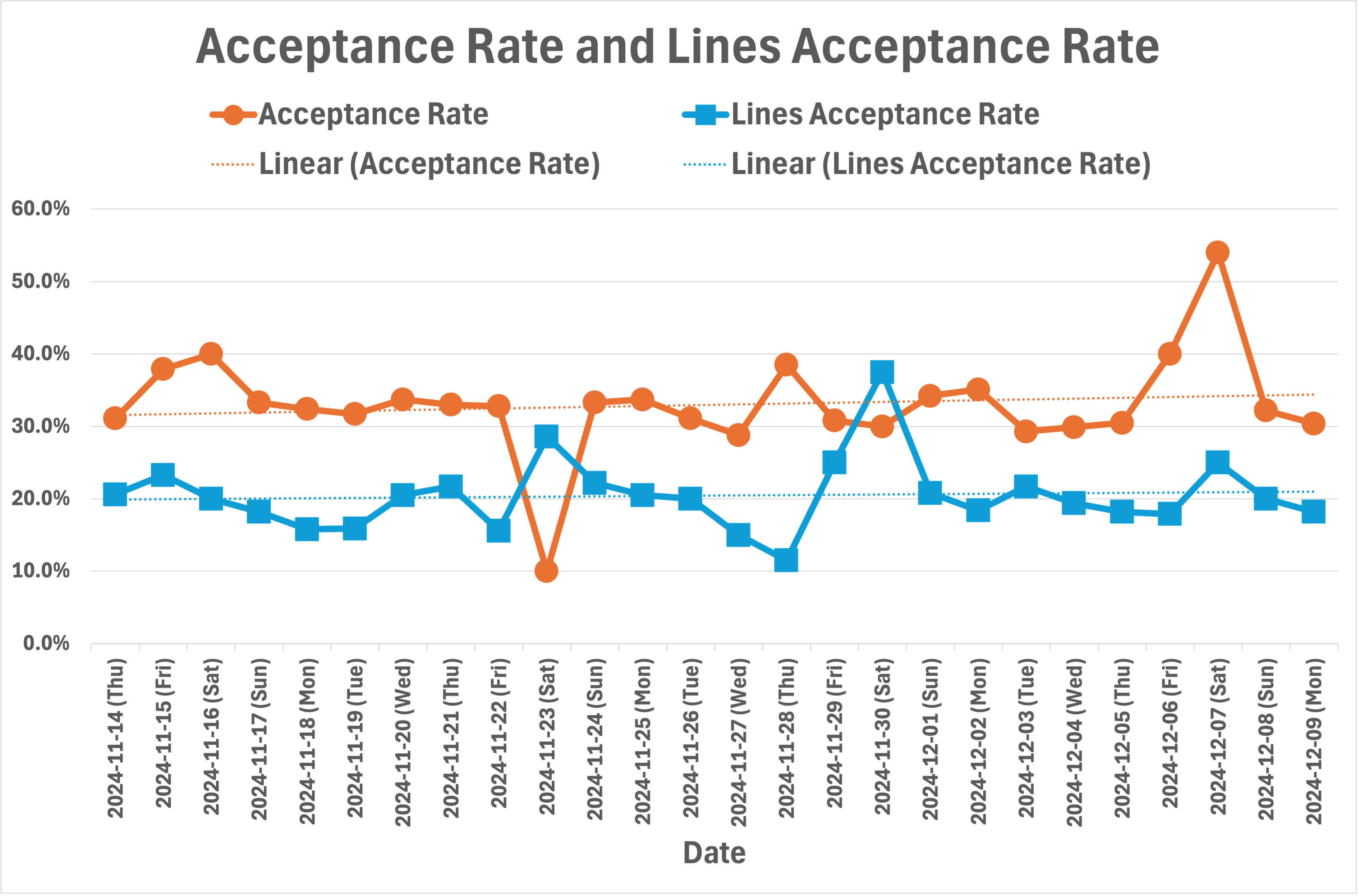}
  \caption{Acceptance rates for suggestions and lines. The former is
    about 1.5 times larger than the latter. The trend lines for each
    also show slight upward trends.}
  \label{fig:rates}
\end{figure}

As shown in Fig.~\ref{fig:rates}, for the period discussed in this
section, the average acceptance rate over suggestions is 33\% and over
lines is 20\%. In other words, one third of the suggestions and one
fifth of the suggested lines are accepted by developers on
average. These numbers are in line with what has been reported by
GitHub~\cite{ZiKaLi24} as well as other companies in industry, e.g.,
\cite{Pi24} while using a different AI pair programmer. It would
interesting to find out in the future why this alignment across
different companies and tools occurs.

When we divide the number of suggestions or lines suggested by the
number of developers, we get a few 10s of suggestions per day per
developer. Though these numbers look reasonably small, the total
number of GitHub Copilot generated lines in our codebase has reached
100s of 1000s of lines. During the time period shown in
Fig.~\ref{fig:table} alone, the total number of lines accepted is
about 75,000. This shows the impact of the tool when used by a large
number of developers.

\begin{figure}[ht]
  \centering
  \includegraphics[scale=0.3]{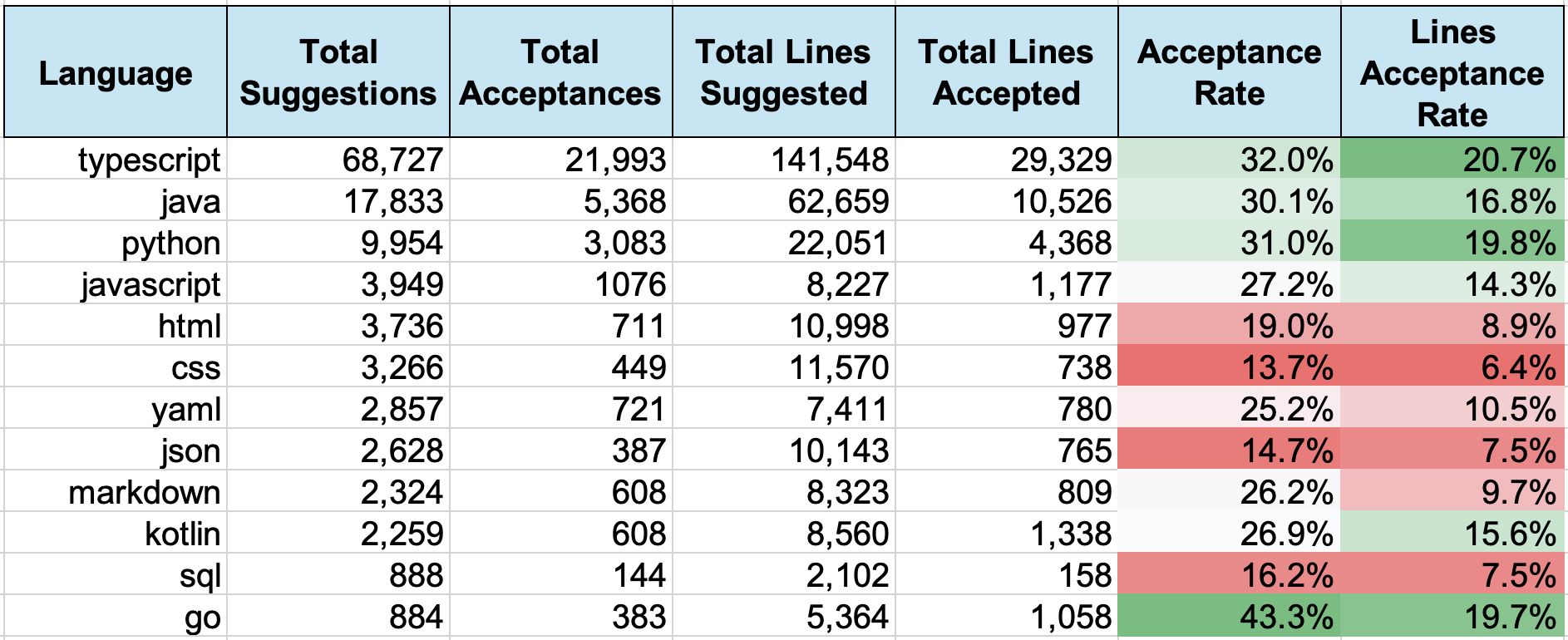}
  \caption{Data for the top dozen languages (collected on Jan. 9th,
    2025. For each language, this table shows the total number of
    suggestions, acceptances, lines suggested, lines accepted,
    acceptance rate, and lines acceptance rate. The conditional
    formatting of two columns in tones of green and red colors are to
    highlight larger values (green) and smaller values (red).}
  \label{fig:lang-table}
\end{figure}

\begin{figure}%
  \centering

  \subfloat[\centering Suggestions per language.]{
    {\includegraphics[width=10cm]{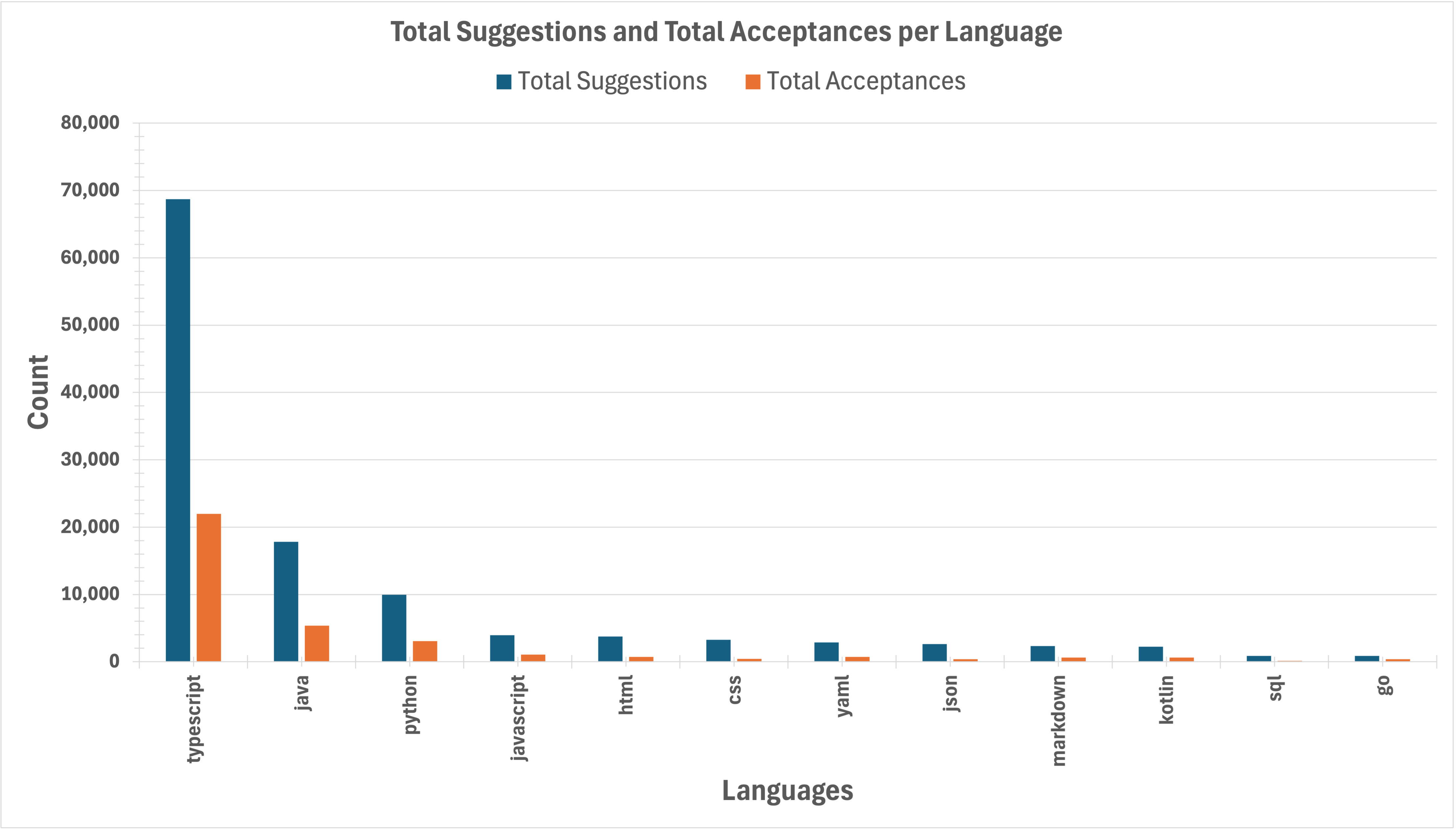} }%
  }

  \subfloat[\centering Lines per language.]{
    {\includegraphics[width=10cm]{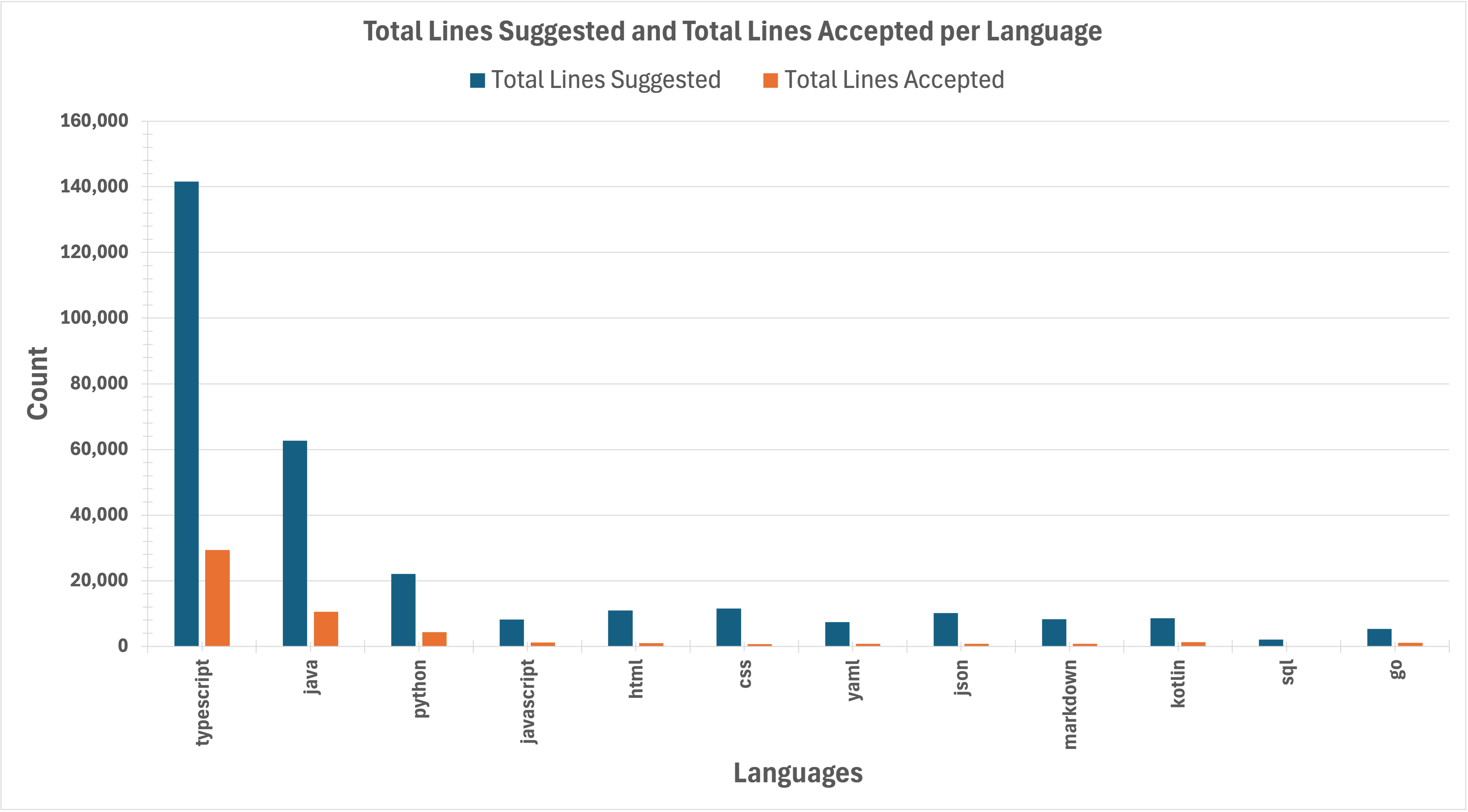} }%
  }
    
  \caption{Daily total number of suggestions and acceptances (a) and
    daily total number of suggested and accepted lines (b) per
    language. Both figures cover the days from Nov. 14th to Dec. 9th
    in 2024. The acceptance rate in each case is the ratio of the
    suggested unit to the accepted unit. The trend lines are for the
    acceptance rates, showing a slight upward trend in each case. The
    wavy pattern is due to weekdays (high) and weekends (low).}%
  \label{fig:lang-vertical}%
\end{figure}

\section{Quantitative Results: Per Language Acceptance Rates}
\label{sec:quantitative2}

Fig.~\ref{fig:lang-table} shows the table of data for about a month in
2024, from from Nov. 11th to Dec. 9th. The table shows per
(programming) language the number of suggestions, acceptances, lines
suggested, and lines accepted. These numbers are the largest for the
top four languages in the table, i.e., TypeScript, Java, Python, and
JavaScript, which is not surprising given the most of our codebase is
in these four languages.

For simplicity, we include data in the table for only the top dozen
languages, sorted in decreasing order of the total number of
suggestions. We excluded data for languages such as Groovy, Shell
(e.g., zsh or bash), Scala, Ruby, and the like since their numbers are
too small.

Fig.~\ref{fig:lang-vertical} shows the same data in the table in
Fig.~\ref{fig:lang-table} but as bar charts and line plots for ease of
understanding. The top four languages cover close to 80\% and 75\% of
the total number of suggestions and lines suggested, respectively, as
well as close to 85\% of the total number of acceptances and lines
accepted.

\begin{figure}[ht]
  \centering
  \includegraphics[scale=0.3]{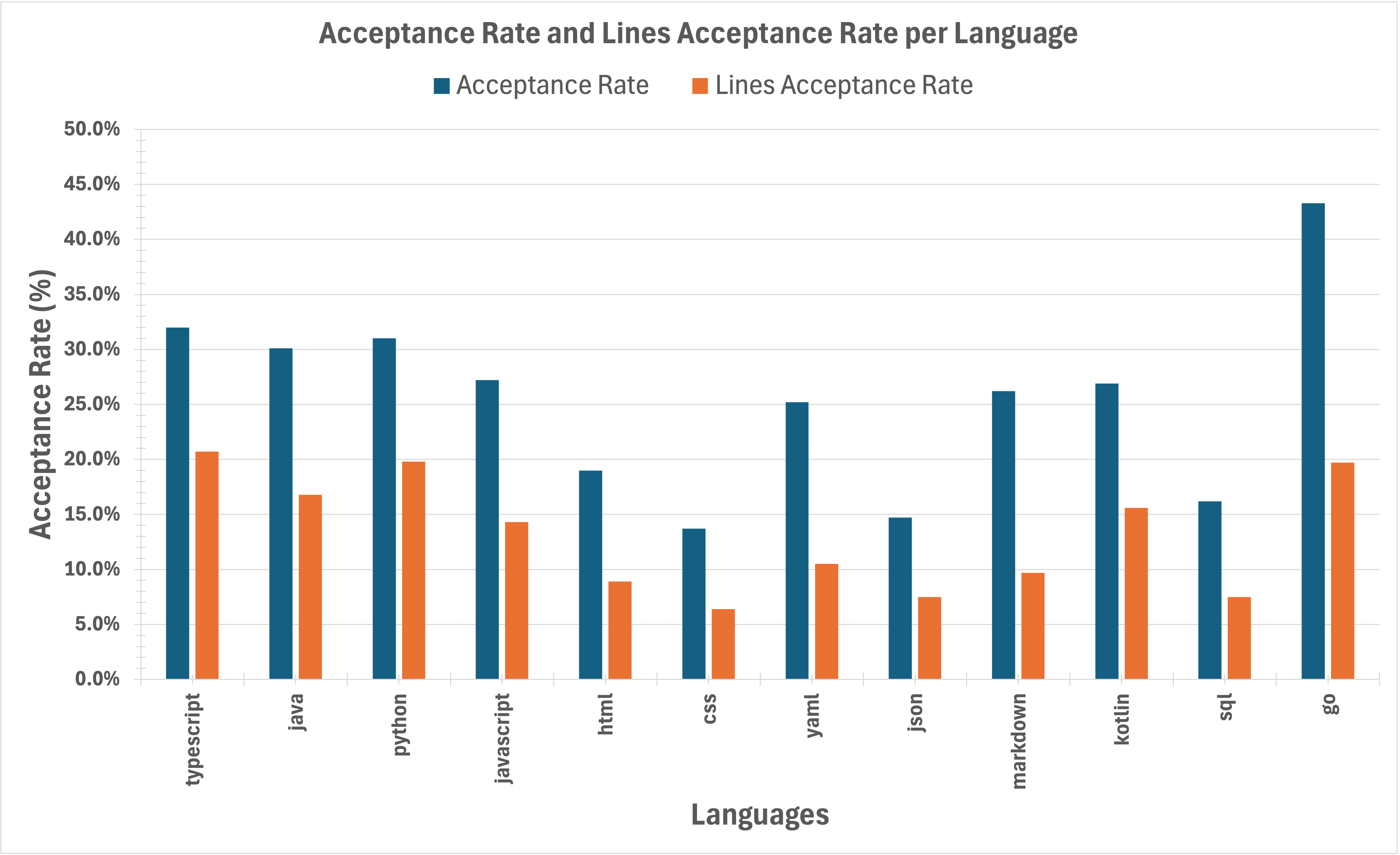}
  \caption{Acceptance rates for suggestions and lines per
    language. The former is 1.5 to over 2 times larger than the
    latter. The trend lines for each also show slight upward trends.}
  \label{fig:lang-rates}
\end{figure}

As shown in Fig.~\ref{fig:lang-rates}, for the 26-day period discussed
in this section, the acceptance rates per language varies between
about 14\% to 32\%. The highest acceptance rate is for the Go language
but the number of suggestions is far smaller. Interestingly the number
of lines per suggestion for the Go language is also the highest, over
6, while the number ranges from 2 to 3 for the other languages.

The acceptance rates for the top three languages are among the
highest, over 30\%, which agrees with the overall acceptance rate we
presented in the previous section. The acceptance rates for HTML, CSS,
JSON, and SQL are interestingly smaller compare to those of the
general-purpose languages. We do not know the reason for this
difference.

\begin{figure}[ht]
  \centering
  \includegraphics[scale=0.3]{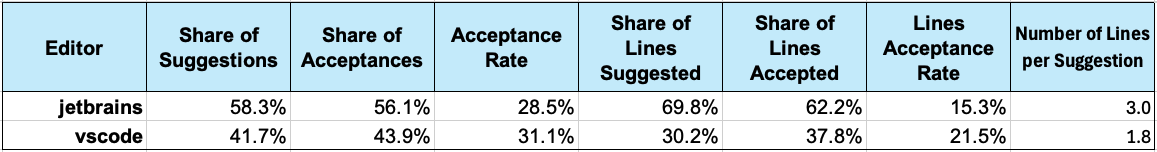}
  \caption{Data for the top two IDEs (``editors'') used by our
    developers, JetBrains and VS Code, in the order of usage. For each
    editor, we show the acceptance rates and share of each editor
    across multiple measures.}
  \label{fig:editors}
\end{figure}

\section{Quantitative Results: Per Editor Acceptance Rates}
\label{sec:quantitative3}

Fig.~\ref{fig:editors} shows the table of data for the top two IDEs
(``editors'') used by our developers: JetBrains (by JetBrains) and VS
Code (by Microsoft). The usage of the former editor is more
widespread, hence, the larger number of suggestions and lines
suggested for JetBrains.

The acceptance rates of suggestions are close to each other and also
close to the 30\% figure we cited for the overall rate. At the same
time, the lines acceptance rates differ with VS Code having about 50\%
higher value. We do not know the reason for this difference but it is
interesting to note the lower number of lines per suggestion for VS
Code.

\begin{figure}[ht]
  \centering
  \includegraphics[scale=0.2]{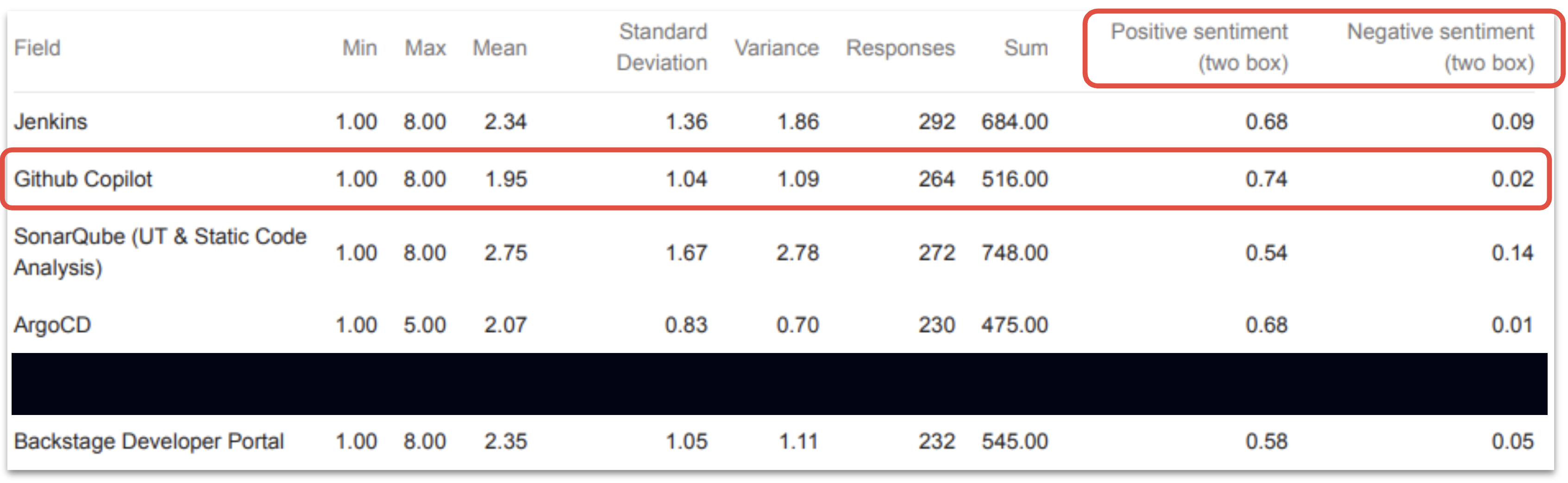}
  \caption{Developer satisfaction statistics. The red boxes indicate
    the relevant parts. The table also shows the data for some other
    tools such as Jenkins, SonarQube, ArgoCD, and Backstage that
    developers love to use. GitHub Copilot has the highest positive
    sentiment. The dark black box hides an internal tool that is not
    relevant to this paper.}
  \label{fig:devsat}
\end{figure}

\section{Qualitative Results: Developer Satisfaction}
\label{sec:qualitative}

In our pursuit of creating the best environment for engineering talent
to thrive, we are committed to employing both quantitative and
qualitative approaches to understand the drivers of developer
productivity. 

We collect qualitative data through quarterly pulse surveys. Our aim
for doing so is to gauge the sentiments and perceptions of our
engineers regarding our development and release toolchain involved in
producing software.

This approach is not only providing structured insights into specific
areas but also allow unstructured feedback to capture the sentiments
and opinions of our engineering workforce. 

By leveraging feedback from our surveys, we are navigating towards
optimizing our investments and identifying areas that require
enhancement. After these surveys are done, we promptly evaluate the
results and communicate back to our developers our learnings and
action items we are planning to take the following quarter.

Since the second iteration of this quarterly survey (Q2 2024), we have
started asking a Likert scale~\cite{Li25} question gauging
satisfaction with various tools in the software delivery toolchain.

Each tool is presented alongside a five-category satisfaction scale
ranging from ``Very satisfied'' (highest) to ``Very unsatisfied''
(lowest) that gauges the satisfaction sentiment for that specific
tool. Open comments are also permitted.

When analyzing the results of the toolchain section, the overall
satisfaction score is calculated as the difference between the two
positive sentiment scores and the two negative sentiment scores, where
the total satisfaction is the result of subtracting the negative
sentiment from the positive sentiment while ignoring
neutralities. This method is similar to those metrics used to measure
customer satisfaction of goods and services offered to customers, and
is referred to internally as ``DevSat'', an abbreviation of
``Developer Satisfaction''~\cite{Ba24}.

As seen in Fig.~\ref{fig:devsat}, GitHub Copilot leads the chart with
the highest satisfaction rate, showing consistent results with 72\%
total satisfaction among all surveyed tools.

In addition to the favorable sentiment, we have the following
observations:
\begin{itemize}

\item 90\% respondents stated that GitHub Copilot reduces the amount of time it
takes to complete their tasks with a median reduction of 20\%.

\item 63\% respondents stated using GitHub Copilot allowed them to complete more
tasks per sprint.

\item 77\% respondents stated that the quality of their work was improved
  when using GitHub Copilot.
  
\end{itemize}

The survey also allows participants to enter free-form text for their
feedback. The following three examples show a positive, somewhat
neutral, and negative feedback.
\begin{itemize}

\item (Positive): ``Github Copilot has been a great productivity tool
  after I learned how to leverage it for certain things.''

\item (Neutral) ``Github Copilot is a hit or miss with correct code in
  VSCode. I don't use the prompt and use the code suggestion if it
  makes sense, but that is wrong a lot of the time. If it's repeated
  code that I already wrote example line(s) of code for, then it's
  usually correct and saves time but it's still one line at a time.''

\item (Negative) ``Copilot sometimes use copilot, usually not giving me
  good results.''
\end{itemize}

\section{Limitations: Observed and Potential}
\label{sec:limitations}

These are the limitations of GitHub Copilot that we were able to
observe so far:
\begin{itemize}

\item Contextual Understanding: Struggles with understanding
  domain-specific logic, leading to irrelevant or redundant
  suggestions.

\item Security Concerns: Potential security risks from auto-generated
  code requiring additional vetting.

\item Creativity Limitations: Generates predictable, less innovative
  solutions in some cases.

\end{itemize}

Though we have not observed yet, we can envision a number of potential
limitations with GitHub Copilot, as follows.
\begin{itemize}

\item Sensitive Data Exposure: GitHub Copilot may inadvertently
  suggest or generate code containing sensitive or proprietary
  information due to its training on public repositories.

\item Intellectual Property Issues: GitHub Copilot might suggest code
  snippets resembling copyrighted or proprietary code, leading to
  potential intellectual property infringement.

\item Code Quality and Vulnerabilities: The generated code could
  contain security vulnerabilities, requiring thorough review before
  being integrated.

\item Data Privacy: If telemetry data from GitHub Copilot is
  collected, there could be privacy concerns over how this information
  is stored and used.

\item Compliance Risks: Some industries have strict compliance
  requirements, and GitHub Copilot may unintentionally violate these
  by generating code or comments not in line with regulatory
  guidelines.

\item Over-reliance on AI: Developers might become too reliant on
  AI-generated suggestions, potentially overlooking best practices in
  favor of quicker implementation.

\item Unintended Patterns: GitHub Copilot might reinforce problematic
  or insecure coding patterns it has learned from public repositories.

\item GitHub Copilot helps with writing code but may make developers
  less creative and even less productive eventually. It is important
  to think about how developers should use it and not rely on it too
  much. Developers should try to balance working fast with coming up
  with their own ideas.

\end{itemize}

\section{Related Work}
\label{sec:related}

There are many factors that affect developer productivity. A
high-level summary of those factors is given in
\cite{Da24}. \S~\ref{sec:productivity} presents a preview. For this
section, we will consider only the factors that can be impacted by
GitHub Copilot. For more information, refer to
\cite{Do24a,Do24b,FoStMa21,NoStFo23,Ta24}.

The news on the launch of GitHub Copilot is given in \cite{Fr21} with
some additional information in \cite{Wi24}. The tool is launched as
``a new AI pair programmer that helps you [developers] write better
code.'' The description of the tool reads ``GitHub Copilot draws
context from the code you’re working on, suggesting whole lines or
entire functions. It helps you quickly discover alternative ways to
solve problems, write tests, and explore new APIs without having to
tediously tailor a search for answers on the internet. As you type, it
adapts to the way you write code—to help you complete your work
faster.''

GitHub Copilot uses a version of Codex, which is originally a GPT
language model from Open AI fine-tuned on publicly available code from
GitHub~\cite{ChTwJu21}. The Codex introduction in \cite{ChTwJu21}
discusses limitations and impact of the model and provides related
work on similar tools. The limitations and impact mention
over-reliance, misalignment, bias and representation, economic and
labor market impacts, security implications, environmental impacts,
and legal implications. The paper also touches upon some risk
mitigation.

There are a growing number of publications on measuring the impact of
GitHub Copilot on developer productivity.

A comprehensive, somewhat ``official'', report on the productivity
impact of the tool is presented in \cite{ZiKaLi24} by researchers from
GitHub, the company that had created the tool. The researchers use
survey responses from developers using the tool as well as
measurements collected from IDEs (interactive development environment)
used by the same developers. The measurements include the number of
suggestions, the number of acceptances, and the amount of code
contributed by and accepted by the developers. The researchers ``find
that acceptance rate of shown suggestions is a better predictor of
perceived productivity than the alternative measures.'' As we discuss
in \S~\ref{sec:success}, we have adopted the same measure to gauge the
productivity impact of the tool. For more details on the measurements
and other aspects of the tool, refer to \cite{Gi24}.

According to \cite{NgNa22}, correctness from the code suggested by
GitHub Copilot is around 60\% for Java and around 30\% for
JavaScript. The paper uses problems from a popular interview questions
site, LeetCode, to generate solutions and then runs LeetCode's problem
tests to evaluate correctness.

\cite{SoAgWe24} uses open source development hosted at GitHub and
finds that GitHub Copilot enhances project-level productivity by 6.5\%
as measured by code contributions; the enhancement is due to increases
in both individual productivity and participation. On the negative
side, the paper finds increase in integration time by about 42\%.

\cite{ZhLiZh23c} collects programming-related discussions about GitHub
Copilot from two popular sites, Stack Overflow and GitHub. The
analysis shows that JavaScript and Python are the most commonly used
languages, Visual Studio Code is the most used IDE, and data
processing is the most commonly used function for code generation. The
discussions also point out challenges with integration of generated
code, confirming the related findings in \cite{SoAgWe24}.

\cite{HaBrZa24} evaluates GitHub Copilot for test generation and finds
that the tool is not a panacea. It uses tests generated by the tool
for open source projects. Within an existing test suite, the paper
finds that about 55\% of generated tests fail while outside an
existing test suite, about 92\% of generated tests fail.

\cite{YeOzTu22} does a systematic evaluation for limits and benefits
of GitHub Copilot; it evaluates the quality of the generated code and
finds that valid code is generated with around 92\% success
rate. Diving deeper, it also finds that suggested code is correct at
around 29\%, incorrect at around 20\%, and partially correct for the
rest.

\cite{MaPaGu23} evaluates the robustness of code generation using
GitHub Copilot. The study prompts the model using semantically
equivalent natural language descriptions and expects the same
suggested function as the result. In close to 50\% of the cases, the
generated function is expected to be the same but is
different. Moreover, the correctness also gets affected in about 30\%
of the cases. The findings shed some negative light on the robustness
of the tool while highlighting the importance of prompting and
experimenting.

\cite{We23} uses two versions of the Codex, one behind GitHub Copilot
and another (Davinci) that is the most performant, to run experiments
on simple programming assignments to measure the correctness of the
generated code and the limitations of the tool. The paper reports
somewhat negative results on Copilot's capabilities compared to
Davinci; it also emphasizes that students using the tool need to be
aware of its limitations; the onus is on them to correct the generated
code for the final solutions.

\cite{PuSp22} is similar to \cite{We23} in that both study GitHub
Copilot in tasks for introductory programming assignments. This paper
reports more positive results for the tool, stating that the generated
code correctness ranges from around 70\% to 95\% when evaluated by
human graders.

\cite{DaMaNi22} studies the capabilities of GitHub Copilot on
generating solutions for fundamental algorithmic problems like sorting
and basic data structures implementations; it also studies the
performance with respect to human programmers on a number of programming
tasks. The paper reports that while Copilot generates correct
solutions most of the time, its performance is still not as good as
human programmers.

\cite{ChLiRo24} is a work similar to ours in many respects: deployment
and use on real-world engineering tasks in a corporate environment,
four weeks of evaluation, use by about 1,000 engineers, and results
assessment via surveys and objective data collection using controlled
experiments. The paper reports notable boost in productivity, code
quality, and job satisfaction due to the use of the tool. The study
measures productivity using the time spent solving a given programming
tasks. The boost is approximately between 40\% and 50\% with the gap
in average time spent between those engineers not using the tool and
those using it growing in favor of the latter as the difficulty of the
task is increasing.

Taking the impact of GitHub Copilot on individual productivity as a
given, \cite{YeMaOe24} evaluates the impact on a collaborative work
setting by focusing on collaborations to open source projects. The
paper finds more maintenance-related contributions than
code-development contributions.

GitHub Copilot is not the only AI model for developer
productivity. For example, AI models for bug fixing and unit test
generation are presented in \cite{TuWaBa19} and \cite{TuDrSv20},
respectively. Both models learn from all the code hosted at
GitHub. Both papers claim positive impact on developer productivity.

There are multiple other studies comparing GitHub Copilot against
similar tools.

\cite{YeOzAy23} compares GitHub Copilot against two other tools,
Amazon's CodeWhisperer and Open AI's ChatGPT. The paper reports
Copilot taking the second place after ChatGPT in the quality of the
generated code on the HumanEval benchmark problems proposed and also
used by the authors of \cite{ChTwJu21}.

\cite{YuLiHu24} attempts to measure the performance of GitHub Copilot
and ChatGPT on code generated by such tools and hosted at GitHub. The
paper dismisses the use of the HumanEval benchmark problems as not a
representative of the real-world problems. The paper reports that
GitHub Copilot and ChatGPT are the top two models used; Python, Java,
and TypeScript are the most common languages for code generation
especially for data processing and transformation while C/C++ and
JavaScript are the most common ones for algorithm, data structure, and
user interface implementations. It also reports that the generated
code is mostly small functions with low complexity and sparse
comments; the generated code by these tools also goes through fewer
modifications compared to human generated code.

There are a couple of studies about aspects of GitHub Copilot beyond
code correctness alone. 

\cite{KoWaNo24} experiments with three human languages used for
prompting GitHub Copilot: Chinese, English, and Japanese. Possibly not
surprisingly due to the amount of training data in these languages, it
finds the worst performance occurs with Chinese and the performance
for all three languages drops with increasing question difficulty.

\cite{NiMiMa23} introduces and evaluates a semi-automated pipeline for
extracting sensitive personal information from the Codex model used in
GitHub Copilot. It finds that the generated code for about 8\% of test
cases has privacy leaks.

\cite{LaGu23} gathers instructor perspectives about how they plan to
adapt to these AI coding tools that more students will likely have
access to in the future. It states that there is no consensus yet; in
the short term instructors are divided on whether to ban or allow AI
usage.

After this related work review, we feel our findings regarding
benefits and limitations of GitHub Copilot are mostly in alignment
with those of the relevant related work but the exact figures
naturally differ due to such reasons as types of tasks, interview
questions vs. production work, programming languages, students
vs. developers, etc. 

\section{Conclusions and Future Work}
\label{sec:final}

In addition to sharing our comprehensive evaluation methodology, our
study provides several key insights into the enterprise-scale
deployment of GitHub Copilot for developers:

\begin{itemize}
\item Quantitative impact: The tool shows consistent acceptance rates
  (33\% for suggestions, 20\% for lines) across different programming
  languages, indicating reliable utility across diverse development
  contexts.
\item Developer satisfaction: High satisfaction rates (72\%) and
  positive feedback suggest that GitHub Copilot effectively supports
  daily development tasks, particularly in areas like boilerplate code
  generation and unit testing.
\item Implementation considerations: Our phased deployment approach,
  including security training and policy alignment, proved effective
  for managing the transition to AI-assisted development.
\item Limitations and challenges: While generally successful, the tool
  shows limitations in understanding domain-specific logic and
  requires careful consideration of security implications.
\end{itemize}

Future work should focus on long-term impact assessment, particularly
regarding DORA metrics, code quality, and maintenance
implications. Additionally, investigating the tool's effect on
developer learning and skill development would provide valuable
insights for enterprise adoption strategies. 

\bibliographystyle{plain}
\bibliography{copilot}
\end{document}